\pdfoutput=1
\documentclass{JINST}

\newcommand{\zctsper}     {{$10^{-2}$~cts/(keV$\cdot$kg$\cdot$yr)}}
\newcommand{\tctsper}     {{$10^{-2}$~cts/(keV$\cdot$kg$\cdot$yr)}}

\newcommand{\dctsper}     {{$10^{-3}$~cts/(keV$\cdot$kg$\cdot$yr)}}

\newcommand{\kgyr}        {{kg$\cdot$yr}}

\newcommand{\cum}         {{m$^3$}}

\def\powten#1{{$10^{#1}$}}

\newcommand{\qbb}         {{$Q_{\beta\beta}$}}

\newcommand{\bbno}        {{$0\nu\beta\beta$}}
\newcommand{\onbb}        {{$0\nu\beta\beta$}}

\newcommand{\up}          {\rule{0mm}{5mm}}

\newcommand{\etal}        {\textit{et al.}}

\newcommand{\gerda}       {\textsc{Gerda}}

\newcommand{\igex}        {\textsc{Igex}}

\newcommand{\hdm}         {\textsc{HdM}}

\newcommand{\gesix}       {{$^{76}$Ge}}

\newcommand{\geenr}       {{$^{\rm enr}$Ge}}          
\newcommand{\genat}       {{$^{\rm nat}$Ge}}
\newcommand{\gedep}       {{$^{\rm dep}$Ge}}
\newcommand{\geox}        {{GeO$_2$}}

\newcommand{\thzza}       {{$^{228}$Th}}

\title{Isotopically modified Ge detectors for 
  {\sc Gerda}:  from production to operation}

\author{D. Budj{\'a}{\v{s}}$^i$, M. Agostini$^i$, L. Baudis$^l$,
    E. Bellotti$^{e,f}$, 
  L. Bezrukov$^g$, R. Brugnera$^{j}$, C. Cattadori$^f$, A. di Vacri$^a$,
  R. Falkenstein$^k$, A. Garfagnini$^{j}$, S. Georgi$^d$, P. Grabmayr$^k$,
  A. Hegai$^k$, S. Hemmer$^{j}$, M. Hult$^c$, J.~Janicsk{\'o} Cs{\'a}thy$^i$,
  V. Kornoukhov$^{g,h}$, B. Lehnert$^b$, A. Lubashevskiy$^d$, S. Nisi$^a$,
  G. Pivato$^{j}$, S. Sch{\"o}nert$^i$, M. Tarka$^l$, and K. von Sturm$^k$\\
\llap{$^a$}  INFN Laboratori Nazionali del Gran Sasso LNGS, Assergi, Italy\\
\llap{$^b$}  Institut f{\"u}r Kern- und Teilchenphysik
             Technische Universit{\"a}t Dresden, Dresden, Germany\\
\llap{$^c$} Institute for Reference Materials and Measurements, Geel, Belgium\\
\llap{$^d$} Max Planck Institut f{\"u}r Kernphysik, Heidelberg, Germany\\
\llap{$^e$}  Universit{\`a} di Milano Bicocca, Milano, Italy\\
\llap{$^f$}  INFN Milano Bicocca, Milano, Italy\\
\llap{$^g$}  Institute for Nuclear Research of the Russian Academy of
            Sciences, Moscow, Russia\\               
\llap{$^h$} Institute for Theoretical and Experimental Physics, Moscow, Russia\\
\llap{$^i$}  Physik Department E15 and Excellence Cluster Universe,
            T.U. M{\"u}nchen, Germany\\                      
\llap{$^j$}  Dipartimento di Fisica e Astronomia dell{`}Universit{\`a} di
Padova  e INFN, Padova, Italy\\
\llap{$^k$}  Physikalisches Institut, Eberhard Karls Universit{\"a}t
            T{\"u}bingen, T{\"u}bingen, Germany\\
\llap{$^l$}  Physik Institut der Universit{\"a}t Z{\"u}rich, Z{\"u}rich,
            Switzerland\\
  E-mail: \email{grabmayr@uni-tuebingen.de}}

\abstract{
 The \gerda\ experiment searches for the neutrinoless double beta (\bbno)
 decay of \gesix\ using high-purity germanium detectors made of material
 enriched in \gesix. For Phase~II of the experiment a sensitivity for the half
 life  $T_{1/2}^{0\nu}\,\,\sim2\cdot10^{26}$~yr is
 envisioned. Modified Broad Energy Germanium detectors (BEGe) with thick n$^+$
 electrodes provide the capability to efficiently identify and reject
 background events, while keeping a large acceptance for the \bbno-decay
 signal through novel pulse-shape discrimination (PSD) techniques.
 The viability of producing thick-window BEGe-type detectors for the
 \gerda\ experiment is demonstrated by testing all the production steps from
 the procurement of isotopically modified germanium up to working BEGe
 detectors.  Comprehensive testing of the spectroscopic as well as PSD
 performance of the \gerda\ Phase~II prototype BEGe detectors proved that the
 properties of these detectors are identical to those produced previously from
 natural germanium material following the standard production line of the
 manufacturer.

 Furthermore, the production of BEGe detectors from a limited amount of
 isotopically modified germanium served to optimize the production, in
 order to maximize the overall detector mass yield. The results of this test
 campaign provided direct input for the subsequent production of the enriched
 germanium detectors.}


\begin{document}

\section{Introduction}
\label{sec:intro}
 \gerda~\cite{Gerda,gerda_tec} is an experiment located at Laboratori
 Nazionali del Gran Sasso (LNGS) of INFN that searches for the neutrinoless
 double beta (\bbno) decay employing high-purity germanium (HPGe) detectors
 enriched in the isotope \gesix. The bare detectors are operated in liquid
 argon in order to minimize the background from radioactive isotopes in
 materials near the detectors. The first phase of the experiment is employing
 reprocessed detectors (18~kg total mass, enriched to 86\% in \gesix) from the
 past experiments \hdm\ and \igex~\cite{HdM,IGEX}.  \gerda\ aims at a
 reduction of the background in the region of interest at \qbb= 2039~keV to a
 level of \tctsper, which is about a factor of 10 lower than previous
 state-of-art. So far, a background index (BI) of $\sim$2$\cdot$\zctsper\ has
 been reached~\cite{gerda_tec}. In the absence of a signal and given the
 current BI, \gerda\ expects to set 90~\% probability lower limits of $T_{1/2}
 > 1.9\cdot$\powten{25}~yr for an exposure of 20~kg$\cdot$yr in
 Phase~I~\cite{gerda_tec} to scrutinize the claim~\cite{hvkkclaim}.  In order
 to improve the half-life sensitivity by a factor~10 in Phase~II of the
 experiment, 20~kg of new detectors from enriched germanium (\geenr) will be
 added to reach the proposed exposure of 100~\kgyr\ within reasonable time.
 Simultaneously, the BI is expected to be reduced to \dctsper\ or
 below~\cite{Gerda}.

 Apart from improved shielding, active background suppression techniques will
 be employed to reach the envisioned BI of $\stackrel{<}{\sim}$\dctsper. The
 measures comprise the instrumentation of the liquid argon and new enriched
 germanium detectors.  The chosen detector type for Phase~II is a modified
 thick window Broad Energy Germanium (BEGe) detector manufactured by Canberra
 Olen, Belgium~\cite{Olen}. Their proper functioning in LAr has been
 demonstrated~\cite{BEGeLAr}. In comparison with the presently used coaxial
 germanium detectors, BEGe detectors have lower noise which translates into
 better energy resolution. Furthermore, their characteristic internal electric
 field distribution allows the identification of \bbno-decay like signals and
 rejection of background signals with higher sensitivity and robustness using
 a novel technique of pulse-shape discrimination (PSD)~\cite{BEGePSA,BEGePSS}.

 Compared to coaxial HPGe detectors used in \gerda\ Phase~I and past
 \onbb\ experiments~\cite{HdM,IGEX}, the BEGe detector design imposes tighter
 constraints on the impurity and defect concentrations in the germanium
 crystal material. These requirements arise due to the non-coaxial electrode
 arrangement in BEGe detectors and the resulting electric field profile; these
 are more demanding to achieve complete charge collection within the detector
 volume.

 To demonstrate that working BEGe detectors with the desired properties can be
 produced from the procured \geenr\ material, a comprehensive validation was
 performed. The tests of the production chain started from procurement of
 isotopically modified germanium material of the same chemical history as the
 already available \geenr, i.e.  the leftovers from the enrichment process,
 depleted in \gesix, that is named \gedep. Further steps included the
 purification, crystal pulling and diode production. The tests finished with
 spectroscopic and PSD performance characterizations of the manufactured
 detectors.

 A second goal was the optimization of the production processes in order to
 maximize the mass yield of functional detectors from the available amount of
 \geenr\ material. These improvements were directly implemented into the
 production process of the \gerda\ Phase~II \geenr\ detectors. The data
 obtained with the characterization of the \gedep\ BEGes provide an important
 reference for assessing the performance of new \geenr\ BEGes.

 The various steps of the production chain are described in the following
 section~\ref{sec:procure}. Section~\ref{sec:testing} reports the results of the
 detector characterization.

\section{Procurement and production}
 \label{sec:procure}

 The supply chain starts with the procurement of isotopically modified
 germanium in the form of GeO$_2$. The isotopic composition is then confirmed
 by different types of material analyses performed at different
 laboratories. The production process continues by reduction of the oxide to
 its metallic form and subsequent purification via zone refinement. This is
 followed by crystal pulling, the determination of electrically active
 impurity concentrations in the crystal volume, and selection and cutting of
 slices useful for detector production. Detectors are fabricated by
 transforming the germanium crystal slices into junctions that are then
 mounted into vacuum cryostats for tests of their properties.

\subsection{Material procurement and isotopic analysis}
   \label{ssec:history}
 In total, 34~kg of {\gedep}O$_2$ material was acquired in early 2009 from ECP
 Zelenogorsk, Russia~\cite{ecp}. This material was a by-product from the
 previous enrichment process for \gerda\ and was chemically processed
 identically as the enriched germanium material. Given the higher abundance of
 \gedep\ after the enrichment process with respect to the \geenr, the cost per
 mass of \gedep\ is substantially lower. The only physical difference of the
 chemical identical material, is the isotopic composition and in particular
 the abundance of the isotope \gesix.

 Several 1~g sized samples of the GeO$_2$ material were used for isotopic
 analysis via inductively-coupled plasma mass spectrometry (ICPMS) and neutron
 activation analysis (NAA). The combined results for the isotopic abundance
 $f_{Ge}$, i.e. the relative number of nuclei, are shown in
 Table~\ref{tab:isotopes}. The standard deviation is dominated by the
 different results from the various methods rather than by the individual
 statistical uncertainties of the measurements.  An isotopic analysis of the
 material after completing the full production chain is envisioned. Such a
 test could detect possible admixtures of natural germanium (\genat) during
 the production processes. In order to maximize the sensitivity for such
 tests, the isotopic abundance of \gesix\ in the procured material was
 required to be $<$1~\% from the beginning.

\begin{table} [ht]
\begin{center}  
\caption{\label{tab:isotopes}
     Isotopic abundance $f_{Ge}$ 
     of the depleted \geox\ samples, averaged over all
     measurements performed at LNGS (ICPMS), INR RAS, Moscow (ICPMS),
     University of T{\"u}bingen (NAA, with sample irradiation at FRM~II Munich)
     and IRMM, Geel, Belgium (NAA, with sample irradiation at SCK$\bullet$CEN,
     Mol, Belgium).
  }
\vspace*{2mm}  
\begin{tabular}{l|ccccc}    
 & & & & &\\[-2mm]   
 isotope & 70 & 72 & 73 & 74 & 76 \\[1mm]      
\hline 
 & & & & &\\[-2mm]     
 $f_{Ge}$  & 0.223(8) & 0.300(4) & 0.083(2) & 0.388(6) & 0.006(2) \\  
\end{tabular}       
\end{center}     
\end{table}

 \subsection{Crystal production}
 \label{ssec:crystals}

 After taking samples for ICPMS the input was 33.84~kg of \geox\ material that
 was reduced to its metallic form and purified via zone refinement at
 \textit{PPM Pure Metals} in Langelsheim, Germany~\cite{ppm}. This process
 yielded 21.49~kg of 6N grade germanium with specific resistivity
 $\rho_{spec}>$50~$\Omega$cm and $\sim$1.37~kg beyond this
 specification. Using the specified abundances and the atomic mass $M_A$=72.44
 the input mass amounts to 23.47~kg metallic germanium. Thus, the purification
 process was performed with an efficiency of 91.6~\%.

 Crystal pulling and detector slice production was performed at Canberra Oak
 Ridge, USA~\cite{oakridge}. The manufacturer modified its production process
 with the objective to maximize the total mass yield of final detectors for a
 fixed amount of starting germanium material. In total four p-type crystals
 were produced from which 14 detector slices were cut with a total mass of
 10.0~kg. This corresponds to a yield of 46.5~\% with respect to the starting
 21.49~kg of 6N depleted germanium material. Given the exploratory goal of
 this production test, also non-standard crystal slice dimensions and shapes
 were tested. Moreover, slices close to the seed-end and tail-end side of the
 crystal were investigated, to test whether working detectors can be produced.
 The number of test and inspection slices, usually 1~mm thick, was minimized.

\begin{figure}[t]
\begin{center}
\includegraphics[width=0.99\textwidth]{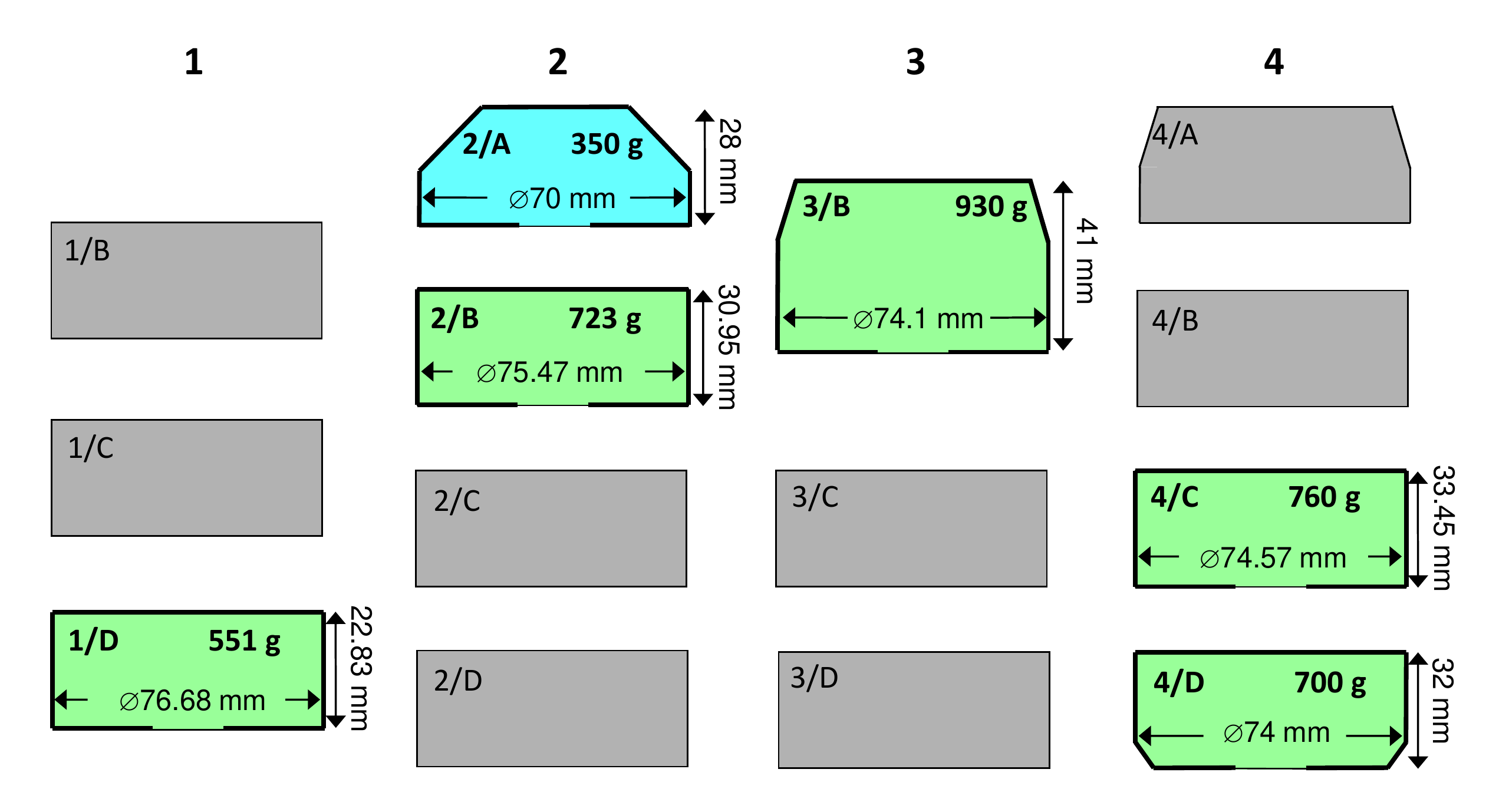}
  \caption{\label{fig:crystals}
          Summary plot of the \gedep\ crystals and their slices. The
          orientation of the crystals is such that the seed end is on the
          top. The slices processed into detectors are shown in green, and the
          one slice which was turned into a functional diode, but not mounted
          into a vacuum cryostat is shown in blue. The thicker border line of
          the processed slices indicates the approximate extent of the
          Li-infused layer. The masses of the slices are given in g.
}
\end{center}
\end{figure}

 Prior to defining the dimension of each detector slice, a full electrical
 field calculation was carried out for the conceived detector geometry,
 redundantly by the manufacturer and by us. For the latter the method of field
 calculations is summarized in Ref.~\cite{BEGePSS}.  Design criteria required
 that the depletion voltage should not exceed 4~kV and that all regions inside
 the detector volume should have sufficient strong electrical fields to ensure
 full charge collection. The diameter of the final slices vary from 72~mm to
 78~mm, the height from 23~mm to 41~mm and the mass from 350~g to 930~g. A
 schematic view of the crystal slices is depicted in
 Fig.~\ref{fig:crystals}. The detector slice identification used in this
 report follows the order of crystal production and a letter corresponding to
 one of maximal four slices cut from each crystal; with A being the seed-end
 slice and D the tail-end slice.

\subsection{Detector manufacturing}
\label{ssec:detectors}

 The detector manufacturing was finalized at Canberra Olen,
 Belgium~\cite{Olen}, by creating the Li-infused n$^+$ electrode covering most
 of the surface, a B-implanted small-area p$^+$ electrode located in the
 middle of the bottom flat face of the cylindrical slice and a circular
 non-conductive groove separating the two electrodes.  The modified BEGe
 detectors for \gerda\ differ from the standard production
 version~\cite{BEGe}.  The latter have a thin entrance window on the top flat
 surface to increase the sensitivity to low energy photons, while the GERDA
 version has a 0.4~mm to 0.8~mm Li-infused n$^+$ electrode also on this part of
 the detector surface.

 The detectors were mounted in standard vacuum cryostats, equipped with
 Canberra 2002CSL charge sensitive preamplifiers. Five crystal slices were
 transformed into detectors with at least one slice taken from each
 crystal. The mass loss turning a crystal slice into a diode is usually of the
 order of ten grams and thus negligible. However, slice 2/A was turned into a
 working diode only after removal of approximately 160~g of material from the
 seed-end side. This diode was not mounted in a vacuum cryostat because of its
 chamfered shape, which is incompatible with standard detector holders.

 The mass yield of crystal slices can be further increased to approximately
 55~\% employing twice the initial mass, as employed in the production of
 enriched detectors. Residual crystal pieces, that are not adequate for
 detector production, can return into the zone refinement or crystal pulling
 process. Moreover, it is estimated that a large fraction of kerf and grinding
 losses can be recovered, chemically reprocessed and returned to the zone
 refinement and crystal pulling process.

\subsection{The time line}
 \label{ssec:time}

 The procurement and production has been achieved in a rather short time.  It
 proved rather helpful that: ($i$) negotiations on buying isotopically
 modified material form ECP started in Moscow on February 26, 2009. The
 material arrived in T{\"u}bingen on April 30. Within one week 40 samples were
 analyzed regarding the isotopic composition by four institutes.  ($ii$) Zone
 refinement at \textit{PPM Pure Metals} was accomplished by June 2009. ($iii$)
 The 6N material arrived at Oak Ridge on August 7. ($iv$) The first shipment
 of three detector grade slices left by plane for Olen on December 16,
 2009. ($v$) The first two detectors were ready on March 19, 2010. ($vi$) The
 second batch of two detector slices was sent to Olen on June 15, 2010.
 ($vii$) The last shipment occurred on January 11, 2011.

 The larger periods between the shipments was owing to intermediate testing at
 Laboratori Nazionali del Gran Sasso. Thus, within less than 2 years the
 production chain could be setup and verified. Having understood the details,
 the enriched BEGe detectors could be produced even faster. The main
 difference concerns a more carefully shielded transport to avoid activation
 of the enriched material.

\section{Detector characterization}
\label{sec:testing}

 The five detectors produced underwent a comprehensive testing campaign to
 characterize their charge collection, spectroscopic and PSD performance, as
 well as long-term stability. An analog data acquisition (DAQ) chain
 consisting of a standard spectroscopy amplifier set to 10~$\mu$s shaping time
 and a multi-channel analyzer was used for the spectroscopic
 measurements. Digital recording of pulse shapes for subsequent off-line PSD
 analysis was accomplished using a DAQ system consisting of a non-shaping
 amplifier and a Struck SIS 3301 14-bit flash analog-digital converter with a
 read-out speed of $10^8$ samples per second similar to the one used for the
 \gerda\ acquisition at LNGS~\cite{gerda_tec}. The recorded pulse-shape data
 were processed and analyzed using GELATIO~\cite{gelatio}, a \gerda\ software
 framework for advanced data analysis and digital signal processing.

 Four out of five detectors show a rather similar behavior resulting in the
 same parameters comparable to the standard detector. Only detector 1/D
 exhibits a reduced performance, the origin of which is not fully clarified.
 It was the last detector delivered and was tested separately. Various
 conjectures will be discussed in the following, e.g. noise contributions or
 inhomogeneous impurity distribution. However, none of these hypotheses can be
 finally rejected or proven.

\subsection{Operational characteristics of the detectors}
\label{ssec:detop}

  The first set of measurements was carried out to investigate the operational
  characteristics of the detectors.  This included the determination of: 
 {\it (i)} the depletion voltage,
 {\it (ii)} the dead layer,
 {\it (iii)} the active volume and other active dimensions, and
 {\it (iv)} the uniformity of charge collection along detector surfaces. 

 The tests were iterated to ensure that a stable optimum for the parameters
 had been reached. Table~\ref{tab:Characteristics} summarizes the main results
 along with the equivalent results for a reference detector -- a BEGe from
 standard Canberra production made of natural germanium, with 81~mm diameter,
 32~mm thickness and a mass of 878~g~\cite{BEGePSA}. The particulars of the
 individual measurements are discussed below. More details about the dead
 layer and active volume measurements can be found in chapter~8 of
 Ref.~\cite{phdTarka}.

 Particularly useful are scans along the surfaces of the detectors.  A
 $^{241}$Am source was collimated to provide a 59~keV photon beam within a
 spot with a diameter of $\sim$1~mm on the detector surface.  Due to the
 collimation, only a small fraction of emitted photons reach the detector. For
 ease of comparison between different detectors and between results and Monte
 Carlo predictions the normalized counts rates $R_n$, namely detected photons
 over emitted photons, are plotted as function of position.  The activity of
 the sources is known to 1.5~\% uncertainty.  The intensity distributions
 $R_n$ will be compared with Monte Carlo (MC) simulations as shown in
 Fig.~\ref{fig:DL-Scan}.

\subsubsection{The depletion voltage}
 \label{sssec:depV}
 The reverse-bias voltage at which the detector volume is maximally depleted
 of free charge carriers, referred to as the depletion voltage $V_{dep}$, was
 determined first. The corresponding volume in the detector is called `active
 volume' $V_{act}$, while the remaining part, in general the surface with the
 `dead layer' and electrodes, remains inactive. The depletion voltage is found
 by irradiating the detector with a $^{60}$Co source while scanning the bias
 high-voltage range and determining the minimal voltage at which the
 spectroscopic characteristics (rate, width and position of $\gamma$ lines)
 are unchanged.  Table~\ref{tab:Characteristics} also includes the operational
 bias voltage $V_{op}$ as recommended by the manufacturer.  $V_{op}$ is a
 conservative value above the full depletion clearing the active volume free
 of charge carriers. The excess of $V_{op}$ over $V_{dep}$ determined by the
 bias voltage scan is typically 1~kV.

\subsubsection{The dead layer}
\label{sssec:deadlayer}
 To determine the thickness of the Li-infused n$^+$ surface dead layer on the
 top area of the detector, the intensities of $^{241}$Am and $^{133}$Ba
 low-energy $\gamma$ lines are compared with Monte Carlo (MC) simulations as
 described in ref.~\cite{phdTarka,phdDB,Optimisation,LNGSBEGe}. The values of
 the measured dead layer, $d_d$, are collected in
 Table~\ref{tab:Characteristics}. The uncertainties include, besides the
 statistical and fit uncertainties, also a measure of the systematic
 uncertainty of the determination methods, expressed as the standard deviation
 of the results obtained with the $^{241}$Am and $^{133}$Ba methods. Knowing
 the thickness $d_d$ the involved mass $m_d$ can be calculated.  A first
 estimate for the active mass fraction from this geometrical consideration can
 made: $f_{geo}= (m_{tot}-m_{d})/m_{tot}$. Values in the range (91 - 96)~\%
 are obtained from this procedure as shown in Table~\ref{tab:Characteristics}.

\subsubsection{The active volume}
\label{sssec:activevol}
 The active volume $V_{act}$ is determined by irradiating the detector with
 $^{60}$Co and reproducing the 1332.5~keV $\gamma$ line detection efficiency
 with MC simulation while assuming that the only inactive part of the detector
 volume is the Li-infused n$^+$ surface dead layer
 (c.f. ref.~\cite{phdTarka,phdDB,Optimisation}). The uncertainty is dominated
 by the uncertainty of the source intensity and of its position with respect
 to the detector. In determining the active fraction of the detector volume
 $f_{av}$, a small contribution to its uncertainty arises also from the
 uncertainties of the outer geometrical dimensions provided by the
 manufacturer.  Comparing the results in Table~\ref{tab:Characteristics}, the
 nice agreement of $f_{av}$ with $f_{geo}$ is observed. Only detector 1/D
 shows a strong deviation that can not be explained so far.

\begin{figure}[t]
\begin{center}
\includegraphics[width=\textwidth]{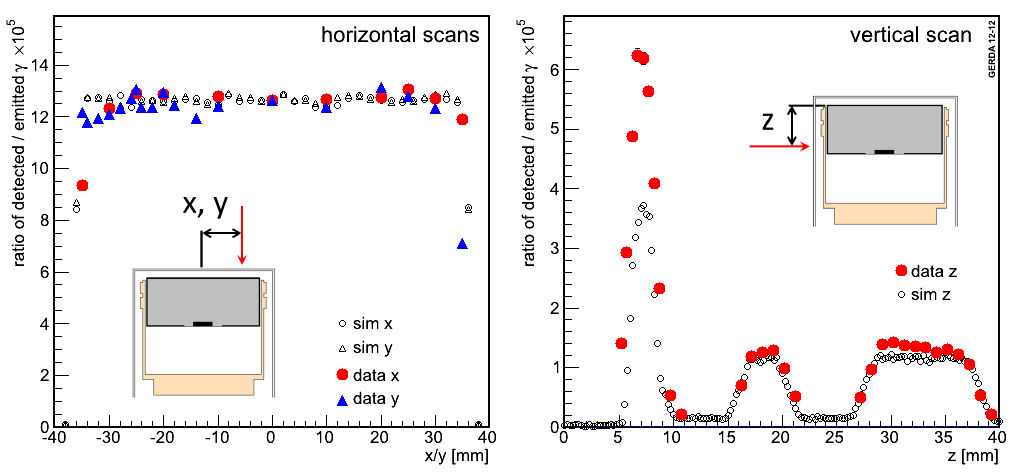}  
  \caption{\label{fig:DL-Scan}
         Normalized intensities $R_n$ for the 59~keV peak of the $^{241}$Am
         source compared to MC predictions. The respective insets show the
         detector vacuum housings and holder cross 
         sections with the beam position and direction indicated.
         Left: $R_{x,y}$ from the horizontal scans in $x$- and $y$- direction
         for the detector 4/D. 
         Right: $R_z$ from a vertical scan along the side surface of detector
         4/C.
}
\end{center}
\end{figure}

\subsubsection{Determination of uniformity from surface scans}
\label{sssec:scanuni}

 The characterization of the uniformity of charge collection along the
 detector surfaces is performed by a scan, that observes the variation of
 spectroscopic parameters (peak position and energy resolution) along the
 scanning points. The variation of these parameters was found to be smaller
 than the random fluctuations caused by electronic disturbances in the DAQ
 setup: typically below $\pm$0.1~keV for the 59.5~keV peak position and below
 $\pm$50~eV for FWHM.

 Secondly, the uniformity of the dead layer thickness was measured by
 observing variation of the peak count rate along the scanning path. The
 uniformity along the top and side surfaces is confirmed within the limits of
 the measurements. This method can provide also an additional measure of the
 absolute thickness of the dead layer. However, the accuracy is much reduced
 compared to the methods described in section~\ref{sssec:deadlayer} due to the
 added uncertainties of the source activity and collimation; the latter being
 dominant in the order of 10~\%.

\subsubsection{Determination of active lengths by surface scans}
\label{sssec:scanactive}

 The surface scans can be evaluated furthermore in order to extract the active
 diameter \o$_{act}$ at the very top and the active height $h_{act}$. From
 both parameters an independent measure for the active volume~$V_{act}$ can be
 derived in principle.  However, since some detectors have a conical shape as
 shown in Fig.~\ref{fig:crystals}, the evaluation of $V_{act}$ of these
 detectors can not be accurate.  The active lengths are measured by detecting
 the reduction of the peak count rate at the edges of the active volume in the
 scans. The obtained values reported in Table~\ref{tab:Characteristics} have
 relatively large uncertainties due to the limited positioning accuracy of the
 scanning device and the relatively large size of the scanning steps.

 The intensity distributions $R_x$ and $R_y$ resulting from the scans on the
 top of the diode along the $x$- and $y$-directions, respectively, are
 compared with MC simulations as shown in the left part of
 Fig.~\ref{fig:DL-Scan}.  The rates are low due to collimation and thus
 multiplied by 10$^{5}$ for display. The uniformity of the charge collection
 along the top is confirmed. As can be noticed, only some minor discrepancies
 between the simulations and the measurement results were observed,
 particularly at the edges. These are most likely to be attributed to an
 imperfect alignment of the scanning device, which was accurate to about 1~mm.

 The right part of Fig.~\ref{fig:DL-Scan} displays $R_z$ obtained from the
 scan along the side surface of detector 4/C. The visible variations in count
 rates are due to blocking of the collimated beam by the copper holder of the
 detector, which has a non-uniform shape. The top of the detector protrudes by
 $\sim$1~mm from the copper holder, around z\,=\,6~mm. Due to the $\sim$0.5~mm
 dead layer, the exposed active part of the detector at this location is
 significantly smaller than the $\sim$1~mm diameter of the collimated
 beam. Therefore the count rate at that position is particularly sensitive to
 the vertical position of the detector in the holder. Small inaccuracies in
 the MC model thus lead to the visible discrepancy between the measurements
 and the MC data.

\begin{table} [t]
\begin{center}
\caption{\label{tab:Characteristics}
         Compilation of selected operational characteristics determined for
         the prototype detectors including the operational voltage $V_{op}$
         given by the manufacturer, the depletion voltage $V_{dep}$, dead
         layer thickness $d_d$, active volume fraction calculated from the
         dead layer $f_{geo}$  and $f_{av}$ as determined by the $^{60}$Co
         scan, the active diameter
         \o$_{act}$, and the active height $h_{act}$. For comparison the
         parameters of a reference natural germanium BEGe from the standard
         Canberra production are shown.
}
\vspace*{2mm}
\begin{tabular}{l|cclllcc}
 & & & & & &\\[-4mm]
  detector & $V_{op}$ & $V_{dep}$ &\multicolumn{1}{c}{$d_d$} & \multicolumn{1}{c}{$f_{geo}$} & \multicolumn{1}{c}{$f_{av}$} & \o$_{act}$ & $h_{act}$ \\[1mm]
 crystal/slice &  [kV] & [kV] &  \multicolumn{1}{c}{[mm]} & \multicolumn{1}{c}{[\cum/\cum]} & \multicolumn{1}{c}{[\cum/\cum]} & [mm] &  [mm] \\[1mm]
  \hline
 & & & & & & &\\[-4mm]
  reference 
     & 4.5 & 3.7 & 0.430(23) & 0.9534(25) & 0.949(18) & 80.0(3) & 31.0(4)\\
 1/D & 3.5 & 2.5 & 0.348(15) & 0.9530(20) & 0.874(13) & - & - \\
 2/B & 4.5 & 3.5 & 0.761(15) & 0.9146(16) & 0.911(12) & 74.3(6) & 29.8(4)\\
 3/B & 5.0 & 3.7 & 0.43(4)   & 0.957(4)   & 0.956(15) & 73.5(5) & 39.5(4)\\
 4/C & 4.0 & 3.0 & 0.64(5)   & 0.931(5)   & 0.919(13) & 72.2(9) & 31.8(5)\\
 4/D & 4.5 & 3.5 & 0.394(12) & 0.9544(14) & 0.932(15) & 72.7(9) & 31.7(5)\\[1mm]
\end{tabular}
\end{center}
\end{table}

\subsubsection{Summary of active and dead volume determinations}
\label{sssec:activesummary}

 The results of these measurements are summarized in
 Table~\ref{tab:Characteristics}.  The measurements of the geometry of the
 active volume are in general consistent with the expectation that the full
 volume of the detector is active except for the surface dead
 layer. Furthermore, the determinations of the dead layer using the comparison
 of the rate in the collimated scans with MC simulations were also in
 agreement with the other methods, despite their lower accuracy.

 The detector 1/D has been an anomaly in this respect: the measured surface
 dead layer makes up (4.7$\pm$0.2)~\% of the total detector volume while the
 active fraction $f_{av}$ measured with $^{60}$Co indicates that only
 (87$\pm$1)~\% of the total volume is active; compare also $f_{geo}$. The
 deviation could be an indication of a deficiency in charge collection inside
 the detector volume. Due to the limited accuracy, the scans along detector
 surfaces were inconclusive with regards to a possible dead layer thickness
 variation. Nevertheless, this discrepancy is not large enough to pose a
 serious concern for the detector production for the enriched BEGe detectors
 of \gerda.

 Actually, as a result of these tests the design of the detector holder was
 changed from the double-ring design to a single ring in order to improve the
 analysis of the scanning of the enriched BEGes. Also its position with
 respect to the surface was clearly stated in the specifications.

\subsection{Tests of the spectroscopic performance}
\label{ssec:spec}

 For \onbb\ experiments the energy resolution and the stability of the energy
 scale are of paramount importance for the identification of a peak that is
 expected to be rather weak.  This peak, if it exists, is more accessible if
 the background can be reduced, e.g. by PSD. These arguments have motivated
 the exploration of the BEGe type detectors to substitute the standard
 semi-coaxial detectors.

 The second part of acceptance tests dealed with the characterization
 of the detector performance. This includes the following measurements of:
(i) {\it energy resolution} of the 1332.5~keV gamma line of $^{60}$Co (see
  Fig.~\ref{fig:FWHM}),
(ii) {\it electronic noise} component of the energy resolution (related to the
  detector capacitance) using a pulser fed to the preamplifier test input,
(iii) {\it PSD performance}; determined with a
  $^{228}$Th gamma-ray source, using the method described in Ref.~\cite{BEGePSA}
  and updated in Ref.~\cite{BEGePSS} (Fig.~\ref{fig:PSD}),
(iv) {\it long term stability} of the charge collection; determined by periodic
  measurements with $\gamma$ sources and a pulser over an interval of
  several months.

 The results of this part are summarized in Tables~\ref{tab:Performance}
 and~\ref{tab:PSD}, while the measurements are described in the following
 subsections.

\subsubsection{Energy resolution}
 \label{sssec:eres}

\begin{figure}[t]
\begin{center}
\includegraphics[width=0.99\textwidth]{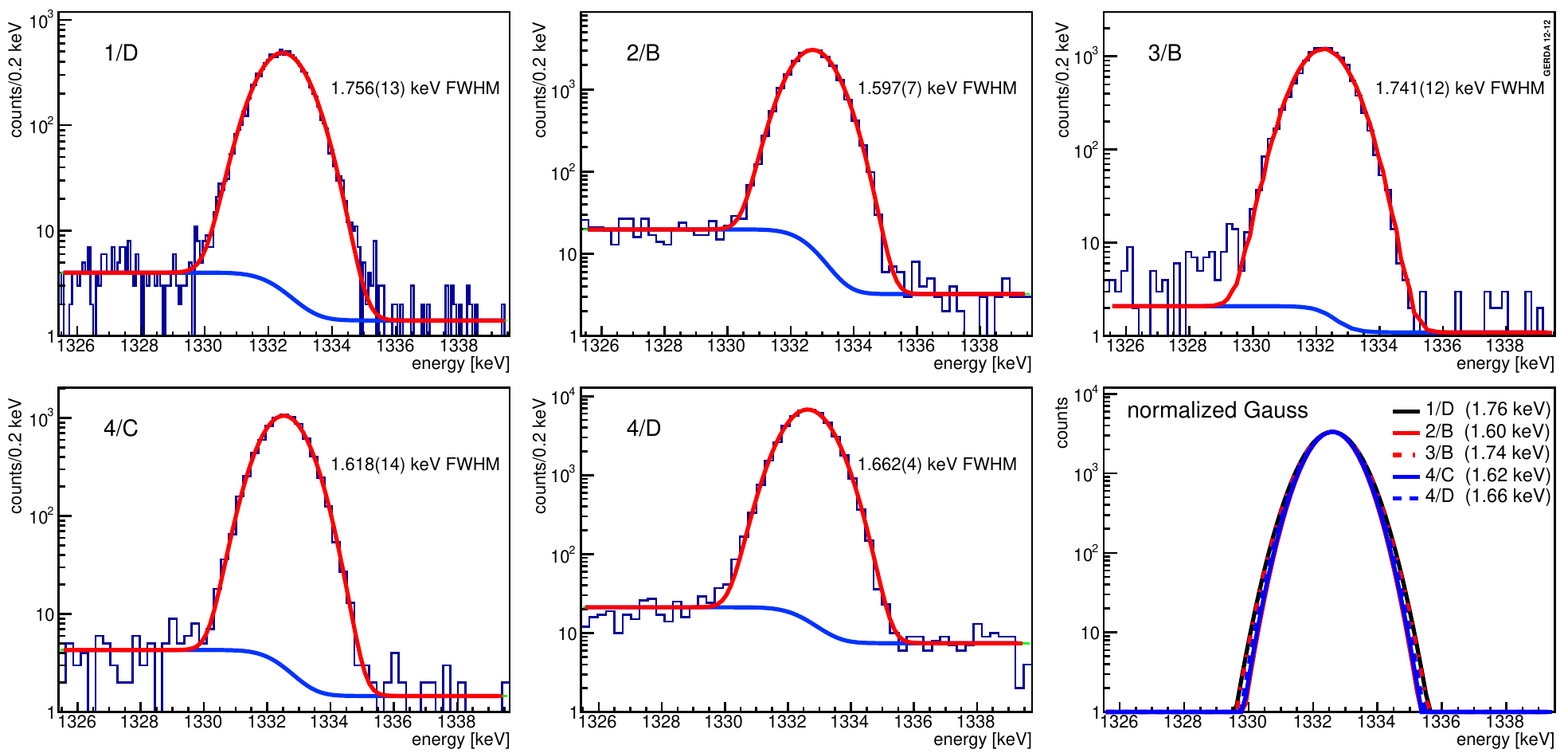}
  \caption{\label{fig:FWHM}
     The responses of the five detectors to the 1332.5~keV $\gamma$ line of
     $^{60}$Co. The spectra are fitted with a step-like background and a
     Gaussian with the FWHM as shown in the plots. The bottom right graph
     shows all Gaussians normalized to the same height for comparison.
}
\end{center}
\end{figure}

 The determination of the energy resolution follows the standard prescriptions
 of spectroscopy, i.e. to accumulate high statistic spectra with the pulser
 and the $^{60}$Co source. The FWHM is the key parameter for the
 comparison. Fig.~\ref{fig:FWHM} shows the zoom onto the peak at
 1332.5~keV. This range is fitted with a step-like background and a Gaussian
 resulting in the widths as shown in the plots.

\begin{table}[b]
\begin{center}
\caption{\label{tab:Performance}
      The energy resolution of the prototype detectors in comparison to the
      reference detector and the duration of the long-term stability tests.
}
\vspace*{2mm}
\begin{tabular}{l|cc|c}
 & & &\\[-2mm]
  detector & pulser & $^{60}$Co & long-term test \\[0mm]
  crystal/slice & FWHM [keV] & FWHM [keV] & duration [d] \\[1mm]
  \hline
 & & &\\[-2mm]
  reference & 0.49 & 1.63 & 51 \\
  1/D & 0.74 & 1.76 &  - \\
  2/B & 0.45 & 1.60 & 32 \\
  3/B & 0.45 & 1.74 & 40 \\
  4/C & 0.43 & 1.62 & 85 \\
  4/D & 0.44 & 1.66 & 45 \\
\end{tabular}
\end{center}
\end{table}

 The size of the steps describing the background contributions varies by a
 factor~2 in these five examples. However, due to different geometries of the
 detectors an interpretation of this factor is not meaningful, even more not
 important as in general this background is small, namely on the \%-level.

 Therefore, the widths of the Gaussian can be extracted rather reliably. For
 comparison, the five Gaussians are normalized to the same height and are
 shown in the sixth frame on the right bottom of Fig.~\ref{fig:FWHM}. The
 uncertainties due to the fit as shown in the plots are only a minor fraction
 of the total uncertainty. The latter is dominated by the stability of the
 energy scale. Therefore, the numbers given in Table~\ref{tab:Performance} are
 reduced in the number of significant digits.  All detectors except the 1/D
 exhibit a comparable FWHM for the pulser. Similarly, 1/D has also the worst
 FWHM for the $^{60}$Co line.

 The worse performance of detector 1/D is not understood. Increased pulser
 FWHM is an indication of higher electronic noise, which could be e.g. caused
 by an increased detector capacitance, an effect that possibly is connected
 with the reduced active volume. However, a possible technical issue with the
 signal read-out electronics could not be ruled out as a source of this
 noise. A problem with the charge collection as implicated in
 section~\ref{sssec:activesummary} would in general manifest itself by an
 appearance of a tail on the side of the $\gamma$ peak; however, this is not
 observed.

\subsubsection{Pulse shape discrimination}
\label{sssec:psd}

 The two $\beta$s from the \onbb\ event have a rather short path length in the
 germanium detector depositing their energy in a rather small volume of a few
 mm$^3$ (`single-site').  In contrast, photons of equivalent energy will
 undergo several scatterings requiring a significantly larger volume
 (`multi-site').  This behavior will be reflected in the time structure of the
 charge signal. To investigate the effect and to define analysis cuts one has
 to retire to $\gamma$ sources of higher energy, e.g. to \thzza\ with its
 2614.5~keV quanta. The double escape peak (DEP) is assumed to exhibit the
 same behavior as the \onbb\ event. The kinetic energy of the
 electron-positron pair is deposited in a similarly small volume while the two
 annihilation quanta escape the active volume of the detector. In contrast,
 for the full energy peak (FEP) typically involves events with multiple
 spatially-separated interactions, either by Compton scatterings followed by
 photo-absorption, or at high energies also via pair production and subsequent
 absorption of both annihilation photons. For detailed explanations and their
 validation see Refs.~\cite{BEGePSA,BEGePSS}.

 The \thzza\ source is rather convenient as it provides a 1592.5~keV line,
 which is the DEP of the 2614.5~keV line in $^{208}$Tl. In its vicinity, the
 FEP of the 1620.6~keV $\gamma$ of $^{212}$Bi appears. Both isotopes are
 $^{228}$Th progenies. Furthermore, both are relatively close to the
 \qbb\ value of 2039~keV. 

\begin{table}[t]
\begin{center}
\caption{\label{tab:PSD}
      PSD performance of the prototype detectors. To facilitate the comparison
      among the several detectors, the cut was set to reproduce a 90~\%
      survival probability for the events in the DEP. The uncertainties
      include statistical as well as systematic contributions. The energies of
      the DEP, SEP, and FEP lines are given in keV together with the Co
      summation peak (SP).
}
\vspace*{2mm}
\begin{tabular}{l|ccccc}
 &  \multicolumn{5}{c}{ }\\[-3mm]
  detector & \multicolumn{5}{c}{surviving fraction after PSD analysis ~[\%]}\\
 \cline{2-6}
\up  [crystal/slice] &  DEP & FEP1 & SEP & FEP2 & $^{60}$Co SP \\
                     &  1592.3& 1620.6& 2103.5&2614.5& 2505.7\\[1mm]
  \hline
 & & & & &\\[-4mm]
  reference
      & 90 & $10.4\pm0.9$  & $5.3\pm0.5$ & ~$ 7.7\pm0.4$ & $0.28\pm0.02$ \\
  1/D & 90 & $14.3\pm0.5$  & $8.7\pm0.5$ & $13.1\pm0.5$  &  -.-\\
  2/B & 90 & $10.3\pm1.4$  & $5.0\pm0.6$ & ~$ 7.4\pm0.6$ & $0.093\pm0.011$ \\
  3/B & 90 & ~$ 7.6\pm0.7$ & $4.3\pm0.4$ & ~$ 5.7\pm0.5$ & $0.090\pm0.012$ \\
  4/C & 90 & ~$ 9.0\pm0.3$ & $4.1\pm0.3$ & ~$ 6.1\pm0.5$ & $0.068\pm0.011$ \\
  4/D & 90 & ~$ 8.5\pm0.6$ & $4.7\pm0.5$ & ~$ 6.4\pm0.5$ & $0.088\pm0.009$
\end{tabular}
\end{center}
\end{table}

\begin{figure}[h!]
\begin{center}
\includegraphics[width=\textwidth]{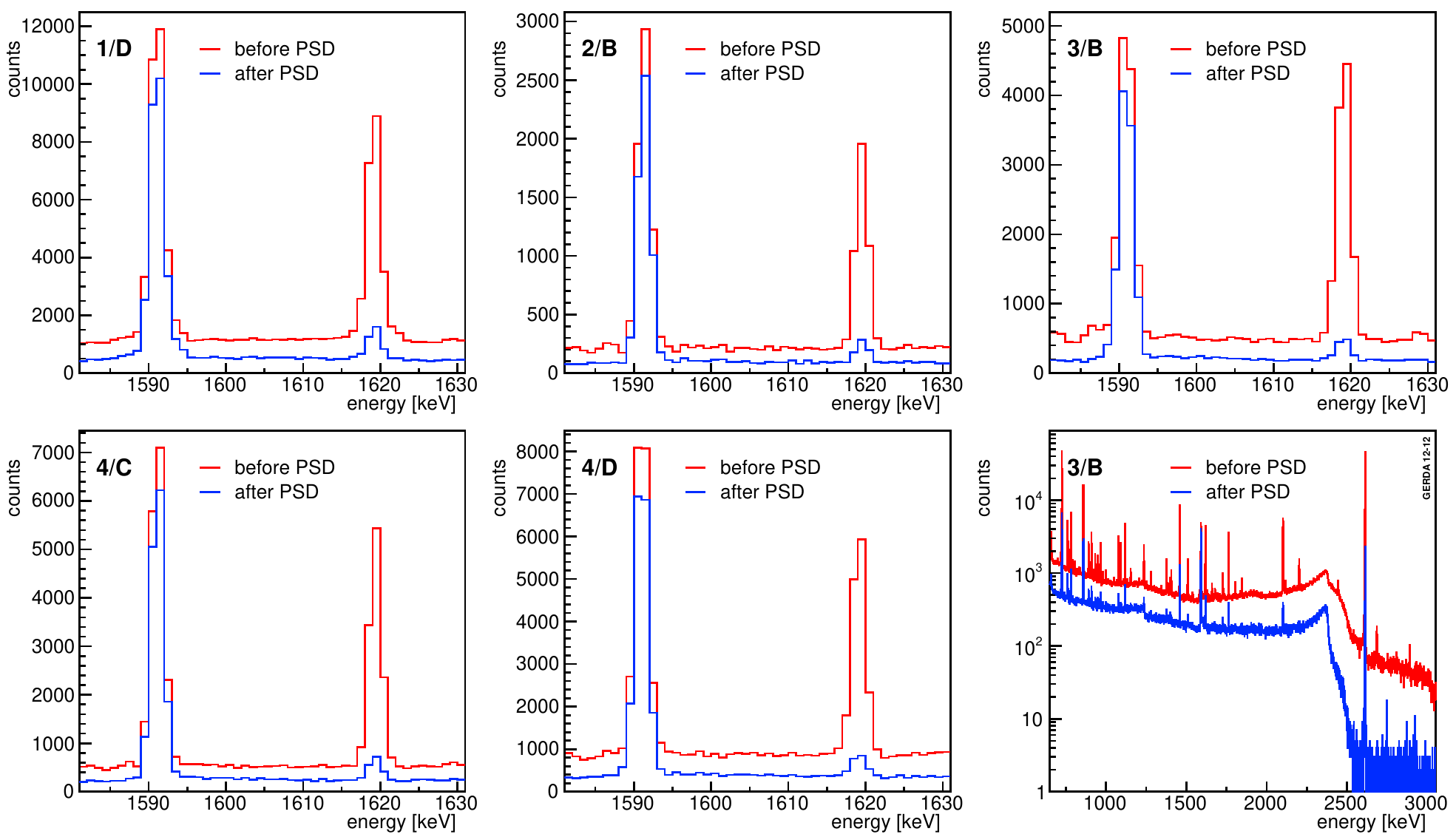}
  \caption{\label{fig:PSD}
    Spectra of $^{228}$Th taken with the five detectors tested, zoomed into the
    range of the DEP at 1592.3~keV and the FEP at 1620.6~keV (see text). The
    standard spectra are given in red (upper curves); applying the PSD cut
    reduces the data as shown in blue (lower curves).  The overall effect of
    the PSD cut is demonstrated by example of detector 3/B for the range from
    650 to 3000 keV (bottom, right, log scale).
}
\end{center}
\end{figure}

 The \thzza\ spectra taken with the five detectors are shown in
 Fig.~\ref{fig:PSD} for the region of the DEP and FEP as discussed above. The
 original spectra are shown as red histograms. Applying the PSD cut after
 optimization and normalizing to a survival fraction of 90~\% for the DEP the
 blue histograms are obtained. In order to appreciate the overall effect of
 PSD a large range from 600 to 3000~keV is plotted in the frame at the bottom
 right of Fig.~\ref{fig:PSD}.

 The effect of PSD is seen clearly in the reduction of the FEP line at
 1620.6~keV for all detectors. The relevant fractions are compiled in
 Tab.~\ref{tab:PSD} where FEP1 corresponds to the discussed line at
 1620.6~keV. Except for the 1/D detector the reduction of the `multi-site'
 events within the peaks to 8--10~\% is reachable. The uncertainties include
 statistical as well as systematic contributions, the latter are derived from
 previous studies as described in Ref.~\cite{BEGePSA,BEGePSS}.  The
 performance is quite similar for all other detectors.

 The bottom right panel in Fig.~\ref{fig:PSD} showing the complete spectrum
 above 650~keV demonstrates the effect of the PSD cut for a wide energy
 range. Actually, it hints at the fraction of `multi-site' events contributing
 to the total spectrum. The separation of SSE and MSE is energy independent
 which was validated in the range from 1.4 to 2.4 MeV, interesting for
 \onbb\ decay.  The reduction is more or less constant at 30~\% up to the
 Compton edge around 2300~keV, only above that energy the intensity is reduced
 to about 5~\% by the PSD cut due to the prevalence of MSE.
 
 Additional information is provided for the single escape peak (SEP) at
 2103.5~keV and the 2614.5~keV line of $^{208}$Tl (FEP2). The last column of
 Tab.~\ref{tab:PSD} shows the PSD performance for the summation peak (SP) of
 the two $^{60}$Co lines that is seen at an energy of 2505.7~keV. The
 summation effect results in highly `multi-site' events, responsible for more
 than two orders of magnitude stronger suppression than is reached in the FEP
 and SEP. The suppression achieved with the four tested detectors is in this
 case much better than that quoted for the reference BEGe, as the analysis
 procedure has ben improved meanwhile.

 The PSD performance of the detector 1/D was again anomalous with higher than
 typical background survival fractions. The cause could possibly be due to the
 higher electronic noise present in the measurements of this detector, or due
 to less favorable electric field distribution inside the detector, possibly
 related to the charge collection deficiency indicated in
 section~\ref{sssec:activesummary}.

\subsubsection{Long-term stability}
  \label{sssec:longterm}

\begin{figure}[t]
\begin{center}
\includegraphics[width=0.99\textwidth]{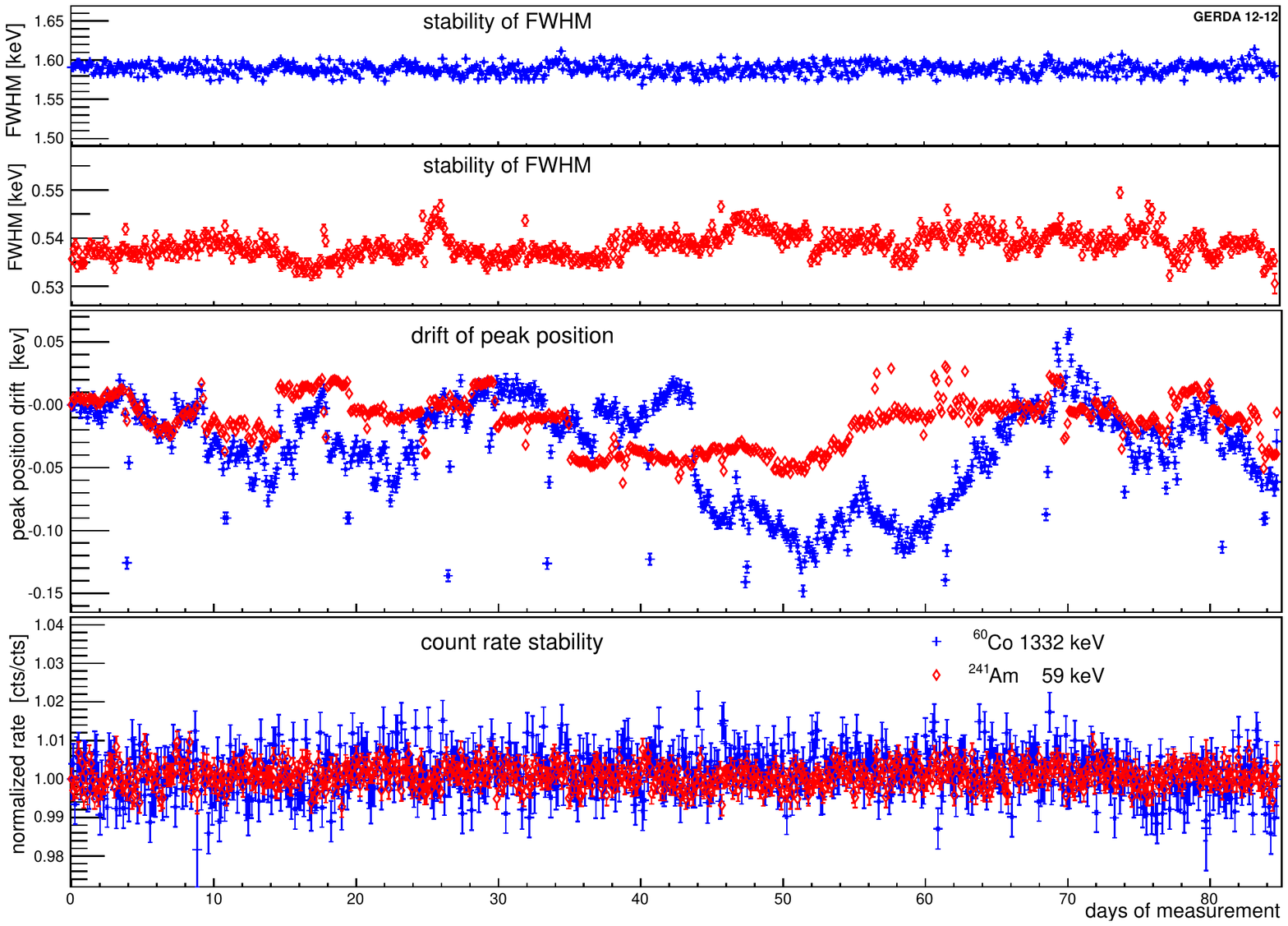}
  \caption{\label{fig:rate}
         Long term stability test of detector 4/C with two sources. The blue
         crosses represent the $^{60}$Co data whereas the red diamonds show the
         results for $^{241}$Am.
}
\end{center}
\end{figure}

 The last column of Table~\ref{tab:Performance} gives the duration of the long
 term tests lasting between 1 and 3 months.  The diodes were irradiated
 alternating with an $^{241}$Am and a $^{60}$Co source. The count rates were
 normalized to the start rate and plotted in Fig.~\ref{fig:rate} together with
 the stability of peak position and energy FWHM. The count rate stays well
 within $\pm$1~\%. Measurement of possible charge collection effects on peak
 position variation was limited by the inherent variability of the DAQ system
 gain (characterized via shorter pulser stability measurements) to the upper
 limit of about $\pm$0.01~\% ($\pm$0.1~keV) at the 1332.5~keV line and about
 $\pm$0.05~\% ($\pm$0.05~keV) at the 59.5~keV line. The FWHM remained also
 very stable within approximately $\pm$20~eV. Similar results were obtained
 with the other detectors, except of 1/D, for which a long term test was not
 performed.

\section{Conclusions}
\label{sec:conclusions}

 The whole production chain, from material procurement to detector operation,
 was successfully tested with depleted germanium left over after the
 enrichment process. Various detectors were produced with modifications of the
 production processes with the purpose to maximize the yield of the upcoming
 new BEGe detectors from the available enriched Ge. The latter, after
 reduction and purification at \textit{PPM Pure Metals}, amounts to 35.4~kg
 plus a 1.1~kg tail of slightly lower than 6N quality which can be recycled in
 a subsequent production run.  The production of new BEGe detectors for the
 second phase of \gerda\ and their characterization is ongoing.

 The measured characteristics and performance of the tested depleted detectors
 were as good as those of a reference BEGe from standard Canberra production
 made of natural germanium. This proves that high-purity germanium material
 crystals produced from isotopic modified germanium raw material according to
 the described production steps have the same properties as those made from
 standard material and processing by Canberra. The depleted germanium has the
 same chemical history as the enriched germanium for the second phase of
 \gerda. Thus, this test confirms the viability of the production chain for
 enriched BEGe detector production.

 Good-performance detectors were produced successfully from germanium crystal
 slices which are normally discarded before BEGe production. In addition, the
 surfaces of some slices were machined less than usual with no loss of
 performance. Thus, the smaller material losses will increase the yield of
 total detector mass from the limited available enriched germanium. Presently,
 the yield of detector mass was about 50~\% with respect to the 6N starting
 material. It can be further augmented by increasing the initially available
 enriched germanium, as in the case of the subsequent enriched germanium
 production. The residual material, including seed and tail ends of crystals,
 as well as kerf and grinding losses can be recovered and chemically
 reprocessed, and a large fraction can return into the crystal production
 process.

 The viability of the production processes of isotopically modified BEGe
 detectors for the \onbb\ decay experiment \gerda\ was proven within a rather
 short period of less than 2 years including all steps. The vital part for the
 enriched program amounted to about 60~\% of this time.

 A precise positioning system for the radioactive sources with well known
 intensity is of paramount importance for the characterization process of new
 detectors. Such a system was established for the tests of the enriched
 detectors. For a \onbb\ experiment the precision of the active volume is the
 parameter of highest relevance followed by the energy resolution itself.

\section*{Acknowledgments}
This work is supported financially by
   the German Federal Ministry for Education and Research (BMBF),
   the German Research Foundation (DFG) via the Excellence Cluster Universe,
   the Italian Istituto Nazionale di Fisica Nucleare (INFN),
   the Max Planck Society (MPG),
   the Swiss National Science Foundation (SNF).
 The institutions acknowledge also internal financial support.

 The authors acknowledge the excellent cooperation with Canberra Oak Ridge,
 Canberra Meriden, and Canberra Olen.

 We thank Thomas Kihm for the DAQ software~\cite{MIZZI} and Matthias
 Laubenstein for valuable help.



\begin{thebibliography}{00}

\bibitem{Gerda}         Letter of Intent to LNGS, see also: hep-ex/04040390,
                        GERDA Proposal to LNGS (2004),
                        {\tt http://www.mpi-hd.mpg.de/gerda/}
\bibitem{gerda_tec}
  K.-H. Ackermann \etal, \emph{The Gerda experiment for the search of
    \onbb\ decay in \gesix},  Eur. Physics J. C {\bf73} (2013) 2330.
\bibitem{HdM}
  M. Gunther \etal, \emph{Heidelberg-Moscow $\beta\beta$
  experiment with \gesix: full setup with five detectors}, Phys. Rev. {\bf
    D55} (1997) 54. 

\bibitem{IGEX}
  C.E. Aalseth \etal, \emph{IGEX \gesix\ neutrinoless double-beta decay
    experiment: Prospects for next generation experiments},
  Phys. Rev. D {\bf65} (2002) 092007.

\bibitem{hvkkclaim}
  H.V. Klapdor-Kleingrothaus \etal, \emph{Search for neutrinoless double beta
  decay with enriched \gesix\ in Gran Sasso 1990-2003},
  Phys. Lett. B586 (2004) 198.

\bibitem{Olen}
         Canberra Semiconductor NV, Lammerdries-Oost 25, B-2430 Olen, Belgium.

\bibitem{BEGeLAr}
 M. Barnab\'e Heider,  D.~Budj\'a\v{s}, K.~Gusev  and S.~Sch\"onert,
 \emph{Operation and performance of a bare broad-energy germanium
  detector in liquid argon}, JINST {\bf5} (2010) P10007.

\bibitem{BEGePSA}
  D. Budj\'a\v{s}, M.~Barnab\'e~Heider, O.~Chkvorets, N.~Khanbekov and
  S.~Sch\"onert, \emph{Pulse shape discrimination studies with a
  broad-energy germanium detector for signal identification and background
  suppression in the \gerda\ double beta decay experiment},
  JINST {\bf4} (2009) P10007. 

\bibitem{BEGePSS}
  M. Agostini, C.~A.~Ur, D.~Budj\'a\v{s}, E.~Bellotti, R.~Brugnera,
  C.~M.~Cattadori, A.~di~Vacri, A.~Garfagnini, L.~Pandola and  S.~Sch\"onert,
  \emph{Signal modeling of high-purity Ge detectors with a small read-out
    electrode and application to neutrinoless double beta decay search in
     Ge-76}, JINST {\bf6}  (2011)  P03005.

\bibitem{ecp}
Currently known as Joint Stock
   Company ``Production Association Electrochemical Plant'' (JSC ``PA
   Electrochemical Plant''), uranium enrichment enterprise of the State Atomic
   Energy Corporation ``Rosatom''.
\bibitem{ppm}
 PPM Pure Metals, GmbH, in Langelsheim, Germany.
\bibitem{oakridge}
   Canberra Oak Ridge, USA.

\bibitem{BEGe}
  Canberra Broad Energy Germanium (BEGe) detector,\\
  {\tt http://www.canberra.com/products/485.asp}. 

\bibitem{gelatio}
  M. Agostini, L.~Pandola, P.~Zavarise and O.~Volynets,
  \emph{GELATIO: a general framework for modular digital analysis
    of high-purity Ge detector signals},  JINST {\bf6} (2011) P08013.

\bibitem{phdTarka}
  M. Tarka, \emph{Studies of Neutron Flux Suppression from
  a $\gamma$-ray Source and The GERDA Calibration System}, Dissertation,
  University of Z\"urich (2012).  

\bibitem{phdDB}
  D. Budj\'a\v{s}, \emph{Germanium detector studies in the
  framework of the GERDA experiment}, Dissertation, University of Heidelberg
  (2009).

\bibitem{Optimisation}
  D. Budj\'a\v{s}, M.~Heisel, W.~Maneschg and H.~Simgen, \emph{Optimisation of
    the MC-model of a p-type Ge-spectrometer for the purpose of efficiency
    determination}, Appl. Radiat. Isot. {\bf67} (2009) 706.

\bibitem{LNGSBEGe}
 M. Agostini, E.~Bellotti, R.~Brugnera, C.~M.~Cattadori, A.~D'Andragora,
 A.~di~Vacri, A.~Garfagnini, M.~Laubenstein,  L.~Pandola, C.~A.~Ur,
 \emph{Characterization of a broad energy germanium detector
  and application to neutrinoless double beta decay search in $^{76}$Ge},
  JINST {\bf6} (2011) P04005.

\bibitem{MIZZI}
 MIZZI Computer Software GmbH, {\tt www.mizzi-computer.de}.

\end{thebibliography}
\end{document}